\documentclass{elsart}

\usepackage{amsmath,amssymb,epsfig}
\usepackage{latexsym,html}
\usepackage{graphicx}

\newcommand{\spec}              { \ensuremath {\mbox{spec}} }
\def\sign#1{\hbox{\rm \,sign}(#1)}

\def\dt{\hbox{\rm d}t}

\newcommand{\C}                 { \ensuremath {\mathbb{C}}  }
\newcommand{\Z}                 { \ensuremath {\mathbb{C}}  }

\newcommand{\hV}                { \ensuremath {\widehat V} }

\newcommand{\hT}                { \ensuremath {\widehat T} }

\newcommand{\hv}                { \ensuremath {\widehat v} }
\newcommand{\hbeta}             { \ensuremath {\widehat \beta} }

\newcommand{\la}                { \ensuremath {\lambda} }
\newcommand{\eq}[1]{eq.~(\ref{#1})}
\newcommand{\eqs}[3]{Eqs. (\ref{#1},\ref{#2},\ref{#3})}

\newcommand{\emu}{{e}_{\mu}}



\def\oT{\underline{T} }

\def\matrix#1#2{\left[\begin{array}{#1}#2\end{array}\right]}

\begin{document}

\runauthor{Van den Eshof et al.}
\begin{frontmatter}



\title{\vspace{-.6cm}\hfill{\rm\small WUB 02-03}\\
Numerical Methods for the QCD Overlap Operator:%
I. Sign-Function and Error Bounds}


%
\author[Utrecht]{J.\ van den Eshof},
\author[Wupmat]{A.\ Frommer},
\author[Wupphys]{Th. Lippert},
\author[Wupphys]{K. Schilling}, and
\author[Utrecht]{H. A.\ van der Vorst}

\address[Utrecht]{Department of Mathematics, University of Utrecht,
The Netherlands}

\address[Wupmat]{Department of Mathematics, University of Wuppertal,
Germany}

\address[Wupphys]{Department of Physics, University of Wuppertal,
Germany}

\begin{abstract}
The numerical and computational aspects of the overlap formalism in
lattice quantum chromodynamics are extremely demanding due to a
matrix-vector product that involves the sign function of the
hermitian Wilson matrix.  In this paper we investigate several methods
to compute the product of the matrix sign-function with a vector, in 
particular Lanczos based methods and partial fraction expansion methods.
Our
goal is two-fold: we give realistic comparisons between
known methods together with novel approaches and we present error
bounds which allow to guarantee a given accuracy when terminating  
the Lanczos method and the multishift-CG solver, applied within the 
partial fraction expansion methods.
\end{abstract}

\end{frontmatter}


\section{Introduction}

Strongly interacting matter, according to the present standard model
of elementary particle physics, is built from quarks interacting by
gluons.  This {\em hadronic} binding problem is highly relativistic:
for instance the mass of the proton is 934 MeV while its constituents,
the light quarks {\sf up} and {\sf down}, carry renormalized masses of
about 5 MeV only.  The fundamental relativistic gauge theory
describing the strong interactions on the level of quarks and gluons
is quantum chromodynamics (QCD).  Due to asymptotic freedom, the
high-momentum sector of QCD can be treated by perturbative methods.
However, on the energy scale of the hadronic binding problem, the
coupling of QCD becomes large.  As a consequence, perturbative methods
do not apply to.  In a seminal paper of 1974, K.\ Wilson has proposed
to treat QCD numerically on a 4-dimensional space-time lattice. This
idea has been brought to life by M.\ Creutz \cite{Creutz:1984mg}
in the form of a stochastic computer simulation on a 4-dimensional
space-time lattice applied to the static quark-antiquark interaction.
Today, after two decades of research, simulations by lattice-QCD are
considered as the {\it only known way} to {\em solve} QCD {\em ab
initio} without recourse to modelling \cite{Lattice:2001}.

A serious shortcoming of Wilson's original discretization of the
fermionic sector of QCD \cite{Wilson:1975hf} is the fact that it
violates, at finite lattice spacing $a$, the {\em chiral symmetry} of
the continuum QCD Lagrangian \cite{Montvay:1994cy}.  Chiral symmetry,
which holds at vanishing quark mass, is of vital importance for our
understanding of the spectrum of elementary particles.  Unfortunately,
violation of chiral symmetry through discretization gives rise to a
host of lattice artifacts, in particular to additive renormalization
of the bare lattice quark mass of the Wilson-Dirac fermion operator,
\begin{equation}
M = I -\kappa D_W.
\end{equation}
The ``hopping term'' $D_W$ is a non-normal sparse matrix, see
\eq{HOPPING}, coupling nearest neighbours on the 4-dimensional
space-time lattice.  The associated Green's function, $G=M^{-1}$, the
quark propagator, is a basic building block of both the stochastic
simulation and the subsequent computation of hadronic operators.  
From a numerical point of view,
the quantity of interest is $G$ multiplied by a source vector $b$,
which can be computed solving the system of equations
\begin{equation}
M\,x=b,
\label{M}
\end{equation}
with efficient Krylov-subspace procedures
\cite{Vorst:1992vd,Frommer:1994vn,Fischer:1996th}.  Such calculations
represent the bulk of the computational effort in lattice-QCD
simulations.

The recent years have witnessed intensive activity in constructing
chiral fermion formulations on the lattice.  An intriguing approach
has been advanced by Neuberger. His so-called overlap operator \cite{Narayanan:2000qx},
defined through
\begin{equation} 
   D=I - M \cdot (M^{\dagger}M)^{-\frac{1}{2}},
\end{equation}
fulfills the Ginsparg-Wilson relation \cite{Ginsparg:1982bj} that has
been re-discovered by Hasenfratz \cite{Hasenfratz:1998gu}. It has been
shown by L\"uscher that this relation implies a novel lattice version
of chiral symmetry \cite{Luscher:1998du}.

The operation of the Green's function of $D$ on the source
vector $b$ can be recast into the equivalent form
\begin{equation} 
\label{neuberger_eq}
(r \gamma_5 + \sign{Q})x = b \quad (|r| \ge 1),
\end{equation}
with $\gamma_5$ being defined in \eq{GAMMA5}.  The matrix $Q$ is the hermitian
form of the Wilson-Dirac matrix, see \eq{HWD}.

The task in solving (\ref{neuberger_eq}) is two-fold:
\begin{enumerate}
\item An {\em outer} iteration with the matrix $r \gamma_5 + \sign{Q}$
has to be performed. Each step of the outer iteration requires one or two
matrix-vector products of the form $\left(r \gamma_5 +
\sign{Q}\right)\cdot v$.
\item Since $\sign{Q}$ is not given explicitly, one needs an additional 
{\em inner}
iteration for each outer step in order to accurately approximate the
product of $\sign{Q}$ with the generic vector $v$,
\begin{equation} 
\label{sign_eq}
s=  \sign{Q} v.
\end{equation}
\end{enumerate}
This latter problem has been dealt with in a number of papers, using
polynomial approximations
\cite{HeJaLe99,Hernandez:1999gu,HeJaLe00,Bu98,Hernandez:2000iw},
Lanczos based methods \cite{Bor99c,Bor99b,Bor99a,Vor00} and
partial fraction expansion \cite{Neu00,EHKN00,EHN98}.  For an overview
consult Ref.~\cite{Wuppertal:1999}.  From these investigations we know
that the computational effort in dealing with overlap fermions is at
least two orders of magnitude larger than with conventional Wilson
fermions due to the repeated application of the sign-function of $Q$
to a vector $v$.

Therefore, it is of particular importance to improve the convergence
of the inner iteration process, which is the focus of the present
paper: we analyze various known methods and present some novel approaches to
iteratively approximate the matrix-vector product with the generic
vector $v$ and the hermitian Wilson-Dirac fermion matrix $Q$.

A key issue that we are addressing in this paper is to determine
explicit accuracy bounds for each step of the inner iteration.
This point is crucial: so
far, for error control a fixed order of the given polynomial
approximation had to be estimated in advance and/or an additional 
end-control using the identity $\sign{Q}^2=I$ had to be performed. 
In practice, this procedure leads to an overhead of at least a factor
of two.
Furthermore, one might suspect that different stages of the outer
iteration require different accuracy of the inner iteration in order
to obtain an overall efficient method, see Ref.~\cite{BFG00}.


This paper is organized as follows: in Section~\ref{optsec} we start
by setting up the general Krylov subspace framework that is common to
all methods considered in this paper (as well as in the literature
known to us).  Within this framework, a `best-you-can-do' method can
be identified. Though infeasible in practice, this method serves for
theoretical comparison purposes, establishing bounds on the maximum
performance which cannot be exceeded by any of the considered methods.
In Section~\ref{statsec} we shortly describe the Chebyshev approach to
approximate (\ref{sign_eq}). This is a numerically feasible method
which has already been used repeatedly in QCD computations. Here it
will serve as reference for comparison with other approaches, from the
practical point of view.

Section~\ref{lancsec} discusses various practical methods for
approximating (\ref{sign_eq}) with the Lanczos algorithm. We consider
known methods as well as new ones and we give a detailed analysis
explaining the behavior of the various methods.  In addition, within
the Lanczos framework, we present a new and cheaply computable
convergence criterion through which a given accuracy in the
approximation to (\ref{sign_eq}) is guaranteed. In this manner we
answer an important question which as yet seems to have remained open.

In Section~\ref{pfesec} we turn to the partial fraction expansion (PFE)
approach as proposed by Neuberger \cite{Neu00}. Here, (\ref{sign_eq})
is approximated with rational approximations to the sign-function
in combination with the multishift-CG algorithm. After discussing two
different existing rational approximations used by Neuberger \cite{Neu00}
and Edwards et al.\ \cite{EHN98}, we propose a new, {\em best}
rational approximation based on a result by Zolotarev, see
\cite{IDK00}. Our approximation can be computed explicitly, that is
there is no need for running the Remez algorithm. 


As a result, the number of poles in the rational approximation (and
thus the number of shifted systems to be solved concurrently in the
shifted CG method) is substantially reduced compared to Ref.~
\cite{Neu00} and also Ref.~\cite{EHN98}.

Moreover, we propose a modification of the multishift-CG method which
saves computational work through early termination of converged
iterations. Again, we develop a procedure to guarantee a given
accuracy.  Section~\ref{numexpsec} presents the results of numerical
experiments for realistic computations on a $16^4$-lattice performed
on the parallel cluster computer ALiCE at Wuppertal University. These
results indicate that our new partial fraction expansion approach is
the most efficient, at least in the present circumstances where  a
two-pass strategy for the Lanczos based methods is necessary.
 
\section{Krylov Subspace Framework} \label{optsec}

The $k$-th Krylov subspace $K_k(Q,b)$ of the operator $Q$ with respect
to the vector $b$ is the linear space spanned by $b, Qb, Q^2b, \ldots,
Q^{k-1}b$,
\[
K_k(Q,b) \equiv \{ p_{k-1}(Q)b : p_{k-1} \in \Pi_{k-1} \},
\]
with $\Pi_{k-1}$ the space of all polynomials of degree $\leq k-1$.

Krylov subspaces are natural subspaces to approximate vectors of the form
$f(Q)b$ with $f$ defined on the spectrum of $Q$, $\spec(Q)$, with values in $\Z$: If
$\spec(Q)$ is given as $\spec(Q) = \{\lambda_1,\ldots,\lambda_n\}$ and if $p \in
\Pi_{n-1}$ interpolates $f$ at the points $\lambda_1,\ldots,\lambda_n$, then
$f(Q)b = p_{n-1}(Q)b$, which shows
$$
f(Q)b \in K_n(Q,b).
$$
We can get approximations $x_k \in K_k(Q,b)$ to $f(Q)b$ by (explicitly or
implicitly) computing a polynomial $p_{k-1} \in \Pi_{k-1}$ which approximates
$f$ on $\spec(Q)$ and setting
$$
   x_k = p_{k-1}(Q)b.
$$
Note that $K_k(Q,b) \subseteq K_{k+1}(Q,b)$, so building up a basis for the
Krylov subspaces can be done in an incremental manner. 

In our QCD context we are concerned with
situations where $f$ is either the $\hbox{sign}$-function
\begin{equation} \label{sign_def}
f(t) = \sign{t} \equiv \left\{ \begin{array}{cc} +1 & t > 0 \\
                                            -1 & t < 0
                           \end{array}
                   \right.
\end{equation}
or the inverse modulus
\begin{equation} \label{inv_kod_def}
f(t) = \frac{1}{|t|} = \frac{1}{\sqrt{t^2}} \quad (\hbox{for } t \neq 0).
\end{equation}
The following observation is crucial: all methods proposed in the
literature on numerical methods for Neuberger fermions
known to us -- including methods based on rational approximations and
all those in the present paper -- turn out to be Krylov subspace
methods, i.e., they determine (different) approximations $x_k \in
K_k(Q,b)$ for $f(Q)b$ with $f$ from (\ref{sign_def}) or
(\ref{inv_kod_def}). We therefore have a common Krylov subspace
framework to present and analyze these methods.

In this context, when trying to compare different methods, it is certainly
interesting to introduce a `best you can do method' -- even if such a method
is not really feasible from a practical point of view. The `best you can do method'
is the following optimal Krylov subspace method: It computes iterates
$y_k \in K_k(Q,b)$ which are best possible in the sense that they have minimal
$l_2$ distance from $f(Q)b$. This means that $y_k$ is the orthogonal projection
of $f(Q)b$ onto $K_k(Q,b)$ which, given an orthonormal basis $v_1,\ldots,v_k$
of $K_k(Q,b)$ can be computed as
$$
   y_k = \sum_{i=1}^k v_i v_i^{\dagger}f(Q)b.
$$
Note that in order to get the coefficients $v_i^{\dagger} f(Q)b$ we need
$f(Q)b$, the quantity we want to compute. So the above optimal method is not
computationally feasible {\em a priori}. But once we have computed $f(Q)b$
to high accuracy, we can use this `best you can do method' for comparison
purposes {\em a posteriori}: For any computationally practical Krylov subspace method with
iterates $x_k \in K_k(Q,b)$ one has
the inequality 
$$
  \| f(Q)b - x_k\|_2 \geq \| f(Q)b - y_k \|_2,
$$
and if the ratio $\| f(Q)b - x_k\|_2 / \| f(Q)b - y_k \|_2$ is close to $1$, the
practical method is a good one. On the other hand, methods for which
the ratio $\| f(Q)b - x_k\|_2 / \| f(Q)b - y_k \|_2$ becomes large are far from
being optimal.

The methods that are most often used in computations for QCD applications
exploit some polynomial approximation, $s(t)$, for $t^{-1/2}$ and 
take $p(t)= t s(t^2)$. Therefore $p$ is an odd polynomial with $p(0)=0$.
In general this restriction can have severe consequences. 
To see this we use the following lemma.
\begin{lem}
\label{lblem}
Define $b^+ \equiv \frac{1}{2}(I+\sign{Q}) b$ and 
$b^- \equiv \frac{1}{2} (I-\sign{Q}) b$.
Let $ r_k^+$ denote the GMRES (see \cite{SS86})
residual at step $k$ for the equation $ Q x
= b^+$ and similarly, $ r_k^- $ for $Q x = b^- $. Then for any polynomial
$ t \cdot q_k (t)$, yielding the approximation $Q q_k (Q) b $ we have
$$
\|\sign{Q}b - Q q_k(Q) b \|_2^2 \geq \| r_{k+1}^+ \|_2^2 + \| r_{k+1}^- \|_2^2 .
$$
\end{lem}
\begin{pf}
We use that $\frac{1}{2}(I+\sign{Q})$ and $\frac{1}{2}(I-\sign{Q})$ are the 
projections onto the two orthogonal invariant subspaces
spanned by the eigenvectors corresponding to the positive and negative
eigenvalues of $Q$, respectively.
  \begin{eqnarray*}
\|\sign{Q}b - Q q_k (Q) b \|_2^2 & = & \| b^+ - b^- - Q q_k (Q) (b^+ + b^-) \|_2^2 \\
& = & \| b^+ - Q q_k (Q) b^+ \|_2^2 + \| b^- + Q q_k (Q) b^- \|_2^2. 
\end{eqnarray*}
The proof follows by noting that
$$
\| b^+ - Q q_k (Q) b^+ \|_2^2  \geq
\min_{p_{k+1} \in \Pi_{k+1},\\p_{k+1}^{(0)} = 1} \| p_{k+1} (Q) b^+ \|_2^2
 = \| r_{k+1}^+ \|_2^2
$$
and similar for the system corresponding to the negative part.
\end{pf}

To illustrate a possible drawback of a
polynomial form with a zero at zero we assume that $Q$ is positive definite.
We then have $\sign{Q}b = b$
and the optimal polynomial is simply given by a polynomial
of degree zero. 
However, from Lemma \ref{lblem} we see that, if we restrict the approximation to the
class of odd polynomials, we need at least a degree equal to the number of iterations required
by GMRES for solving $Qx=b$. Which is, in general, much more.
In Section \ref{qualsec} we show that this restriction on the approximating
polynomials is not a severe restriction in QCD applications.

\section{The Chebyshev Approach} \label{statsec}

A common way to get an approximation from a Krylov subspace for a
matrix function times a vector is using a suitable polynomial
approximation for this function on some set that at least contains the
spectrum of the matrix. For the $\hbox{sign}$-function of $Q$
different polynomial approximations have been proposed on the set
$[-b,-a]\cup[a,b]$, if all eigenvalues of $Q$ are contained in this
interval.  A conceptually very elegant approach is using Gegenbauer
polynomials, see \cite{Bu98}.

As a reference for the other methods discussed in this paper, we
briefly discuss the use of a Chebyshev series and summarize the key
points here; for details on the theory see for instance
Ref.~\cite{FoP72}.

Assume for the moment that $f$ is to be approximated on the interval $
[-1,1]$.  The Chebyshev polynomials $ T_i(t)\in \Pi_i $ are the
orthonormal polynomials with respect to the inner product
$$
\langle f, g \rangle \equiv \int_{-1}^1 \frac{1}{\sqrt{1-t^2}} f(t) g(t) \dt .
$$
We thus have
$$
\langle T_i, T_j \rangle = \delta_{ij}.
$$
Every function $f$ for which $ \langle f,f \rangle $
exists can be 
expanded into its Chebyshev series
\begin{equation}\label{Chebyser_eq}
f = \sum_{i = 0}^\infty c_i T_i
\enspace \mbox{ with } c_i \equiv \langle f, T_i \rangle .
\end{equation}

Truncating the series at the $k$-th summand gives the polynomial
approximation of degree $k$
\begin{equation} \label{Chebypol_eq}
p_k = \sum_{i=0}^k c_i T_i
\end{equation}
which is `best we can do' in a weighted $L_2$-sense, i.e.\ for all
polynomials $q \in \Pi_k$ we have
$$
\langle p_k - f, p_k - f \rangle \leq \langle q - f, q - f \rangle .
$$
Moreover, $ \lim_{i \to \infty} \langle p_i - f, p_i - f \rangle = 0
$, which means that we have convergence in a weighted $L_2$-sense.  If
$f$ is continuous, we also have convergence at every point, i.e.
$$
\lim_{i \to \infty} p_i (t) = f (t) \mbox{ for all } t \in [-1, 1].
$$

For the approximation of $f$ on a general interval $ [\alpha,\beta]
$, we use the linear transformation $ t \to \frac{1}{\beta-\alpha} (2t
- (\beta+\alpha))$ which maps $ [\alpha,\beta] $ onto $ [-1, 1] $. 
This brings us
back to the situation just described. 

In a practical numerical computation, the integral defining the coefficient
$c_i$ in (\ref{Chebyser_eq}) can be approximated by a quadrature rule
$$
c_i \equiv \frac{2}{k} \sum_{j=1}^k f(t_j) T_i(t_j) \approx
\int_{-1}^1 \frac{1}{\sqrt{1 - t^2}} f (t) T_i (t) \dt,
$$
with $t_j = \cos\left(\pi (j-\frac{1}{2})/k\right)$. Moreover, the
approximating polynomial $ p_k $ from (\ref{Chebypol_eq}) is evaluated
by using the numerically stable and efficient Clenshaw-Curtis relation
\cite[Section~3.2]{FoP72}.

For the Dirac overlap operator, truncated Chebyshev series
approximations have been introduced by Hern\'andez, Jansen and
Lellouch \cite{HeJaLe99,HeJaLe00}. Here, the polynomial approximations
$p_k $ from (\ref{Chebypol_eq}) are computed for the function $ f (t)
= 1/\sqrt{t} $ over an interval $ [a^2, b^2] $, where $\spec(Q)
\subseteq [-b, -a] \cup [a,b]$. The approximation to $\sign{Q}b $ is
then obtained as $ Q \cdot p_k (Q^2) b$ where $p_k(Q^2) b$ is
evaluated using the Clenshaw-Curtis recurrence.  We will use this {\em
Chebyshev approach} as a standard for comparison with the other
methods of this paper.

\section{Methods Based on the Lanczos Reduction} \label{lancsec}

The polynomials constructed by the methods mentioned in the previous
section only depend on the radius of the spectrum, i.e.\ on $a$ and
$b$ with $\spec(Q) \subseteq [-b,-a] \cup [a,b]$. In this section and
the next one we discuss methods that construct polynomials which use
implicitly or explicitly information from the Lanczos reduction. From
a practical point of view, this means that they introduce dynamically
computed parameters in the construction of the polynomial.  Therefore,
they are potentially more efficient with respect to the degree of the
required polynomial (the number of matrix-vector multiplications) but
in general are more expensive to construct (they require inner
products and, in general, require more memory).  This dichotomy is
similar to the one known for iterative methods for linear
systems, e.g. \cite[Section 2.2]{Gre97a}.

\subsection{Lanczos approximations}
\label{lancapproxsec} The Lanczos method (e.g.
\cite{Par98,GoL96}) exploits a three-term recurrence for the
construction of an orthonormal basis $v_1$,$\dots$,$v_{k+1}$ for
the subspace $K_{k+1}(A,b)$ as defined in Section \ref{optsec}.
This algorithm is given in Alg. \ref{lanczosalg}.
\begin{algorithm}

\begin{center}
\begin{tabular}{p{8cm}}
  \begin{tabbing}
(nr)ss\=ijkl\=bbb\=ccc\=ddd\= \kill
{\underline {\bf Input:}} a device to compute $Ax$ and a vector $b$ \\
{\underline {\bf Output:}} an orthonormal matrix $V_{k+1}=[v_1,\dots,v_{k+1}]$
and $T_k$ \\
1.\> $v_1 = b/\|b\|_2$, $\beta_0 = 0$ \\
\> {\bf for} $i=1, 2, \ldots, k$ \\
2.\>\> $v = Av_i - \beta_{i-1} v_{i-1} $ \\
3.\>\> $\alpha_i = v_i^{\dagger} v$\\
4.\>\> $v = v - \alpha_i v_i$ \\
5.\>\> $\beta_i = \|v\|_2$ \\
6.\>\> $v_{i+1} = v/\beta_i$ \\
\end{tabbing}  \end{tabular}
\end{center}
\caption{Lanczos algorithm}
\label{lanczosalg}
\end{algorithm}

The Lanczos algorithm can be expressed in matrix form as
\begin{equation}
\label{lanczosmf}
AV_k = V_k T_k + \beta_k v_{k+1} e_k^{\dagger} = V_k \oT_k
\end{equation}
where $T_k$ is a $k \times k$ tridiagonal matrix containing the
$\alpha_i$'s and $\beta_i$'s computed in the Lanczos iteration,
and $\oT_k$ is $T_k$ appended with the additional row
$\beta_k e_k^{\dagger}$,
$$
T_k = \matrix{ccccc}
{
\alpha_1 & \beta_1 \\
\beta_1 & \alpha_2 & \ddots \\
& \ddots & \ddots & \ddots \\
& & \ddots & \ddots & \beta_{k-1} \\
& & & \beta_{k-1} & \alpha_k \\
} \enspace 
\oT_k = \matrix{ccccc}
{
\alpha_1 & \beta_1 \\
\beta_1 & \alpha_2 & \ddots \\   
& \ddots & \ddots & \ddots \\
& & \ddots & \ddots & \beta_{k-1} \\
& & & \beta_{k-1} & \alpha_k \\    
& & & & \beta_k \\
}\; .
$$
We refer to the eigenvalues of $T_k$ as the {\em Ritz values}
(with respect to the search- and test-space $V_k$).

A well-known technique, see \cite{DGK98,Vor88} and the references
therein, for the approximation of $f(A)b$ -- a matrix function times a
vector -- is to reduce the problem to a matrix function of
the low dimensional matrix $T_k$ as in
\begin{equation}
\label{lancapprox}
f(A)b \approx V_kf(T_k)V_k^\dagger b = V_k f(T_k)e_1 \|b\|_2.
\end{equation}

The idea of using this {\em Lanczos approximation} in
(\ref{lancapprox}) for the overlap operator has been considered
by several authors. We review the different approaches in this
section. Our starting point is the Lanczos algorithm for $Q$ with
starting vector $b$, for which the matrix formulation reads
\begin{equation}
\label{lanczosmfQ}
QV_k = V_k T_k + \beta_k v_{k+1} e_k^{\dagger} =  V_{k+1}\oT_k.
\end{equation}
Bori\c{c}i \cite{Bor99a} applies the Lanczos approximation
in (\ref{lancapprox}) to the function
$f(t)=(t^2)^{-1/2} = |t|^{-1}$. This results in
\begin{equation}
\label{borici1eq}
\sign{Q}b =Q|Q|^{-1}\approx  Q V_k {(T_k^2)}^{-1/2} e_1\|b\|_2 =
V_{k+1} \oT_k {(T_k^2)}^{-1/2} e_1\|b\|_2.
\end{equation}
The first expression is the one originally given in \cite{Bor99a};
we have added the second expression to show that the final
multiplication with $Q$ can be circumvented by exploiting the
vector $v_{k+1}$ in Alg. \ref{lanczosalg}. For more details and a
heuristical stopping condition for this method we refer to
\cite{Bor99a}. 

A different approach is proposed by van der Vorst
\cite{Vor00}. He suggests the following approximation
\begin{equation}
\sign{Q}b \approx V_k \sign{T_k} e_1  \|b\|_2 .
\label{vdveq}
\end{equation}
We note that this approximation is contained in the subspace
$K_k(Q,b)$ instead of $K_{k+1}(Q,b)$ as for (\ref{borici1eq}). At
the end of Section \ref{smoothsec}, we will focus on the
difference between both methods.

The errors, as function of $k$, for both methods show large
oscillations; see, for instance, Ref.~\cite[Figure 3]{Bor99a} and our
discussion in the coming sections. In order to avoid such
oscillations, Bori\c{c}i has introduced an alternative method based on
the Lanczos process for $Q^2$ \cite{Bor99c}. If we denote the
corresponding quantities with hats we obtain
\begin{equation}
Q^2\hV_k = \hV_k \hT_k + \hbeta_{k+1} \hv_{k+1}e_k^{\dagger},
\end{equation}
and the resulting approximation is
\begin{equation}
\label{borici2eq}
\hbox{sign}(Q)b = Q (Q^2)^{-1/2} \approx Q\hV_k \hT_k^{-1/2} e_1 \|b\|_2 .
\end{equation}
Numerical experiments indicate that this method indeed converges smoothly.
However, the subspace that contains the approximation $QK_k(Q^2, b)$ is
only a subset of the subspace $K_{2k}(Q, b)$, but it requires the same
number of matrix vector multiplications (MVs) for its
construction (the number of required inner products is less). The
question is whether (\ref{borici2eq}) requires many more MVs than
(\ref{borici1eq}) and (\ref{vdveq}) in order to attain a comparable
accuracy, see also the discussion following Lemma \ref{lblem}. We
return to this question in Section \ref{qualsec}. In Section
\ref{smoothsec} we consider, for theoretical reasons, another Lanczos
approximation based on Lanczos for $Q$, that does have a smoother
convergence and we will compare the different methods.

\subsection{The idea behind Lanczos approximations}

In this section we assume $f$ to be some sufficiently smooth
function. The Lanczos approximation (\ref{lancapprox}) implicitly
constructs a polynomial $p\in \Pi_{k-1}$ that approximates the
function $f$ such that
\begin{equation}
\label{minerr}
\|f(Q)b - p(Q)b\|_2
\end{equation}
is small. It is known \cite[Theorem 3.3]{Sa92} that this polynomial $p$
interpolates the function $f$ in the Ritz values. To understand
the idea behind (\ref{lancapprox}) we consider the more general
polynomial $p\in \Pi_{k-1}$ that interpolates $f$ in the
(distinct) points $\mu_i$, i.e.
\begin{equation}
\label{intpropeq}
p(\mu_i) = f(\mu_i) \quad \hbox{for } i = 1,\dots,k \ ,
\end{equation}
and we consider how these $\mu_i$ should be chosen such that (\ref{minerr}) is small.
For this polynomial we have the following bounds for (\ref{minerr}):
\begin{lem}
\label{intlemma}
For any set of distinct interpolation points $\{\mu_i: i = 1,\dots,k\}$ and for
any function $f \in \C^k$, let $p \in \Pi_{k-1}$  satisfy (\ref{intpropeq}) and
$q(t) \equiv \prod_{j=1}^k(t-\mu_j)$. Then, if all $\mu_i$ and all eigenvalues of $Q$ are contained
in the interval $[\alpha,\beta]$ we have
\begin{equation}
\label{intboundeq}
\|q(Q)b\|_2 \inf_{t \in[\alpha,\beta]} \left|\frac{f^{(k)}(t)}{k!}\right|  \le \|f(Q)b-p(Q)b\|_2 \le
\|q(Q)b\|_2 \sup_{t \in [\alpha,\beta]} \left|\frac{f^{(k)}(t)}{k!}\right|.
\end{equation}
\end{lem}
\begin{pf}
By a standard result on the approximation error of interpolating polynomials,
see \cite[Thm 2.1.4.1]{StoerBu92}, e.g., there exist values
$\xi_i\in[\alpha,\beta]$ such that
\[
f(\la_i)- p(\la_i) = q(\lambda_i) \frac{f^{(k)}(\xi_i)}{k!}.
\]
Now, let $b$ be represented as $b = \sum_{i=1}^n \gamma_i w_i$, where the $w_i$
are orthogonal and unit length eigenvectors of $Q$ with eigenvalue $\lambda_i$.
This gives
$$
\|f(Q)b-p(Q)b\|_2^2 = \sum_{i=1}^n \gamma_i^2
\left(f(\lambda_i)-p(\lambda_i)\right)^2 =
\sum_{i=1}^n \gamma_i^2 q(\lambda_i)^2 \left(\frac{f^{(k)}(\xi_i)}{k!} \right)^2 .
$$

Bounding this expression results in (\ref{intboundeq}).
\end{pf}
As discussed in Section \ref{optsec}, finding the polynomial that
minimizes (\ref{minerr}) is not really feasible. Nevertheless, if $f$ is
smooth enough, we can expect nearly optimal results by choosing the $\mu_i$'s
such that $\|q(Q)b\|_2$ is as small as possible. The
polynomial that minimizes $\|q(Q)b\|_2$ over all monic polynomials in $\Pi_k$
is known as the {\em Lanczos polynomial} and is
equal to $\pi_k(t) \equiv \det(t I - T_k)$, see \cite{PPV95}.
Hence, we expect nearly optimal results when the $\mu_i$ are
equal to the Ritz values. This gives us some justification for the use
of (\ref{lancapprox}) and explains the often good experience with this method in
practice.

The $\hbox{sign}$-function has a discontinuity in zero and therefore the result
from Lemma \ref{intlemma} cannot be applied. However, we can use that only
the function values in the eigenvalues of $T_k$ and $Q$ are of importance and
we are free to replace the $\hbox{sign}$-function with some function that
has a smooth transition around zero. It can happen in practice that
an eigenvalue of $T_k$ is close to zero even though the eigenvalues of $Q$
are far from zero. In this case the smoothed $\hbox{sign}$-function still can have a quite
steep transition around zero.
In the convergence history of the errors we observe this as a peak in the curve.

In order to achieve smoother convergence for `Lanczos on $Q$' we
consider in the next section the use of alternative interpolation
points, known as {\em harmonic Ritz values}. These are bounded away
from zero.

\subsection{Smooth convergence with Lanczos on $Q$}
\label{smoothsec} The convergence of the methods described by
(\ref{borici1eq}) and (\ref{vdveq}) is rather irregular. In fact,
for (\ref{borici1eq}) the error can be infinitely big for a
certain $k$ if $T_k$ has a very small eigenvalue. For
(\ref{vdveq}) the error never exceeds $2 \|b\|$ (because the constructed
approximations have the same length as $b$). 
Therefore the peaks are bounded for this method.

Krylov subspace methods for linear systems that are based on a
Galerkin condition, like CG, often show a similar convergence
behavior for indefinite problems. In this situation the peaks in
the convergence history can cause instabilities in the linear
solver, see e.g. \cite{SVo96d,Gre97b}. For the Lanczos approximations
(\ref{borici1eq}) and (\ref{vdveq}) of the sign function this
poses no serious problem if the case of zero Ritz values is
properly handled. It only requires that we skip the result for
{\em one} iteration. (It can be shown that a Ritz value close to zero in two
consecutive iterations implies that this Ritz value is close to an eigenvalue and
not a so-called ghost Ritz-value.)

The resulting oscillating behavior led Bori\c{c}i to propose his
alternative algorithm, which we discussed in Section
\ref{lancapproxsec}. We will propose an alternative
for (\ref{vdveq}) that is still based on (\ref{lanczosmfQ}). The
idea is to construct a polynomial $p\in\Pi_{k-1}$ that
interpolates $f$ at the {\em harmonic Ritz values} instead of the
Ritz values. The main reason for this is that the harmonic Ritz
values have the property \cite{PPV95} that
the smallest harmonic Ritz value
(in absolute value) is always outside the interval formed by the
largest negative and the smallest positive eigenvalue of $Q$.
Therefore, the harmonic Ritz values stay away from the discontinuity in
the sign-function.

Our main tool is the following lemma which 
generalizes a result by Saad \cite[Lemma 3.1]{Sa92}.
\begin{lem}
\label{polexactlem}
For all $p \in \Pi_{k-1}$ and all $k$-vectors $z$, we have
\begin{equation}
\label{polexacteq}
V_k p(T_k+ze_k^{\dagger})e_1\|b\|_2 = p(A)b .\end{equation}
\end{lem}
\begin{pf}
We show by induction that (\ref{polexacteq}) is correct for all monomials $t^j$ with $j < k$.
Note that $e_k^{\dagger} T_k^j e_1 = 0$ for $j < k-1$.
$$
V_k (T_k+ze_k^{\dagger})e_1\|b\|_2 = V_k T_k e_1 \|b\|_2
= AV_ke_1 \|b\|_2-\beta_k v_{k+1} e_k^{\dagger}e_1\|b\|_2
= Ab
$$
Now suppose (\ref{polexacteq}) holds for $t^i$ with $i \le j$ and let $j<k-1$,
$$
V_k (T_k+ze_k^{\dagger})^{j+1} e_1\|b\|_2 =
V_k(T_k+ze_k^{\dagger})T_k^j e_1\|b\|_2 = V_k T_k^{j+1} e_1\|b\|_2,
$$
$$
V_k T_k T_k^{j} e_1 \|b\|_2=
  AV_k T_k^j e_1 \|b\|_2+ \beta_k v_{k+1} e_k^{\dagger} T_k^j e_1\|b\|_2 =
A^{j+1} b.
$$
\end{pf}
This lemma has an interesting corollary which 
generalizes another result by Saad \cite[Theorem 3.3]{Sa92}.

\begin{cor}
\label{intcoll}
Let $p \in \Pi_{k-1}$ be the unique polynomial that interpolates $f$ in the
eigenvalues of $T_k+ze_k^{\dagger}$. Then we have
$$
V_k f(T_k+z e_k^{\dagger})e_1 \|b\|_2 = p(A)b.
$$
\end{cor}
\begin{pf}
Decompose $f$ in $p+e$ where $p\in \Pi_{k-1}$ is the unique polynomial that
interpolates $f$ in the eigenvalues of $T_k+z e_k^{\dagger}$.
We have that $f(T_k+z e_k^{\dagger}) = p(T_k+z e_k^{\dagger})$. The proof is concluded
by using Lemma \ref{polexactlem}.
\end{pf}
The {\em harmonic Ritz values} are the reciprocals of the
Ritz values of $A^{-1}$ with respect to the search- and test-space $AV_k$
\cite{PPV95},
or, equivalently, the eigenvalues of
$$
T_k^{-1} \left( T_k^2 + \beta_k^2 e_k e_k^{\dagger} \right) =
T_k + z e_k^{\dagger} \hbox{ with } z = \beta_k^2 T_k^{-1}e_k.
$$
It is important to realize that the harmonic Ritz values are all distinct (e.g. \cite{PPV95})
and therefore, we have from Corollary \ref{intcoll} the following
{\em harmonic Lanczos approximation}
\begin{equation}
\label{harlancapprox}
f(A)b \approx V_k f(T_k + z e_k^{\dagger}) e_1 \|b\|_2 \hbox{ with }
z = \beta_k^2 T_k^{-1}e_k.
\end{equation}
In practical applications the matrix function for the small matrix can be computed
using a non-symmetric spectral decomposition (which exists because
the harmonic Ritz values are distinct). However, due to the inverse of $T$ this can lead to inaccurate
results in case there is a Ritz value close to zero.
It is interesting to observe that if $f(t) = t^{-1}$, then the
Lanczos approximation is identical to the CG approximation from
$V_k$ and the harmonic Lanczos approximation is equal to the
MINRES approximation. (Note that the usual implementation of MINRES 
uses a more stable updating procedure than 
(\ref{harlancapprox}).)

\begin{figure}
\centering
\includegraphics[width=\textwidth]{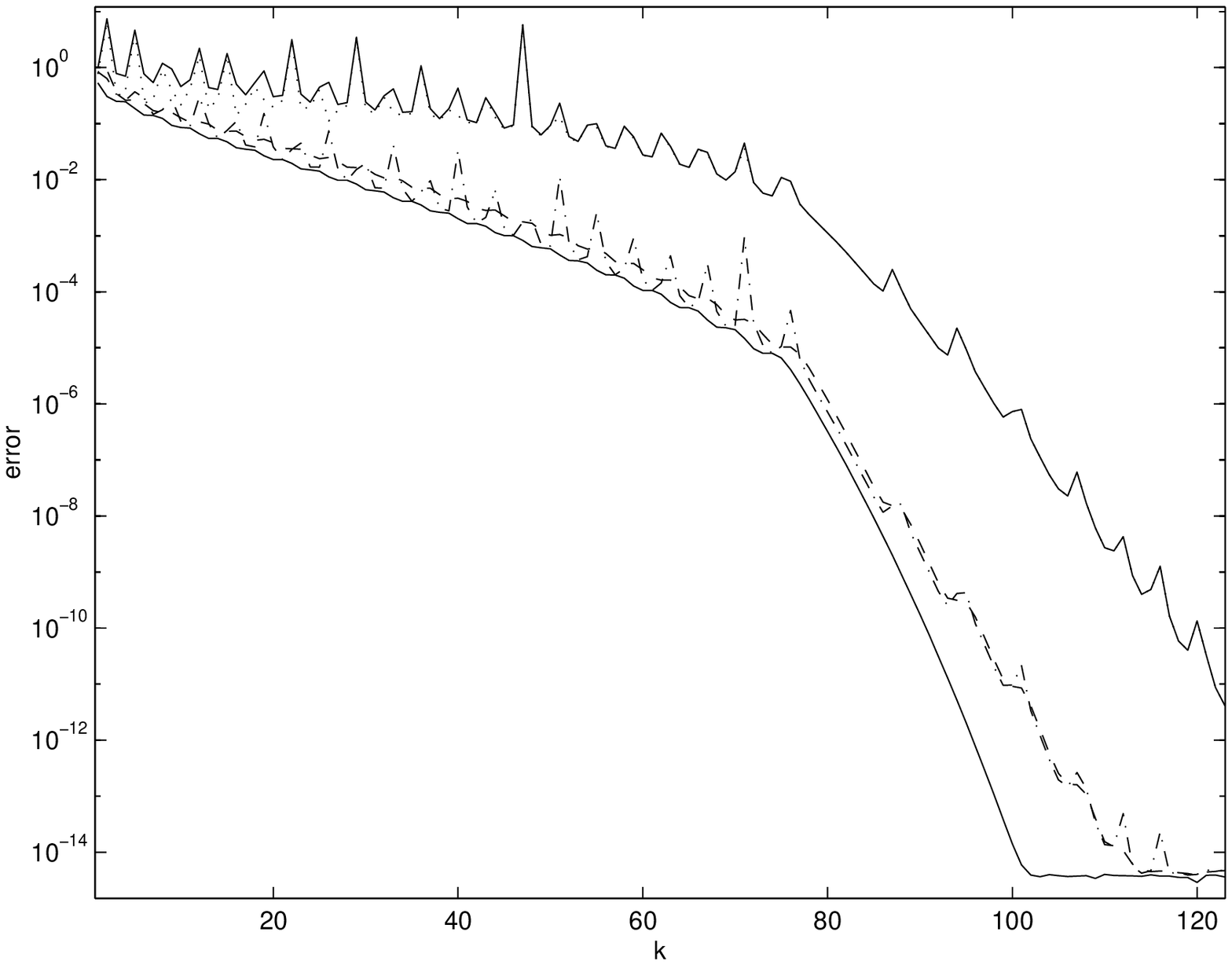}\vspace*{-.5cm}
\includegraphics[width=\textwidth]{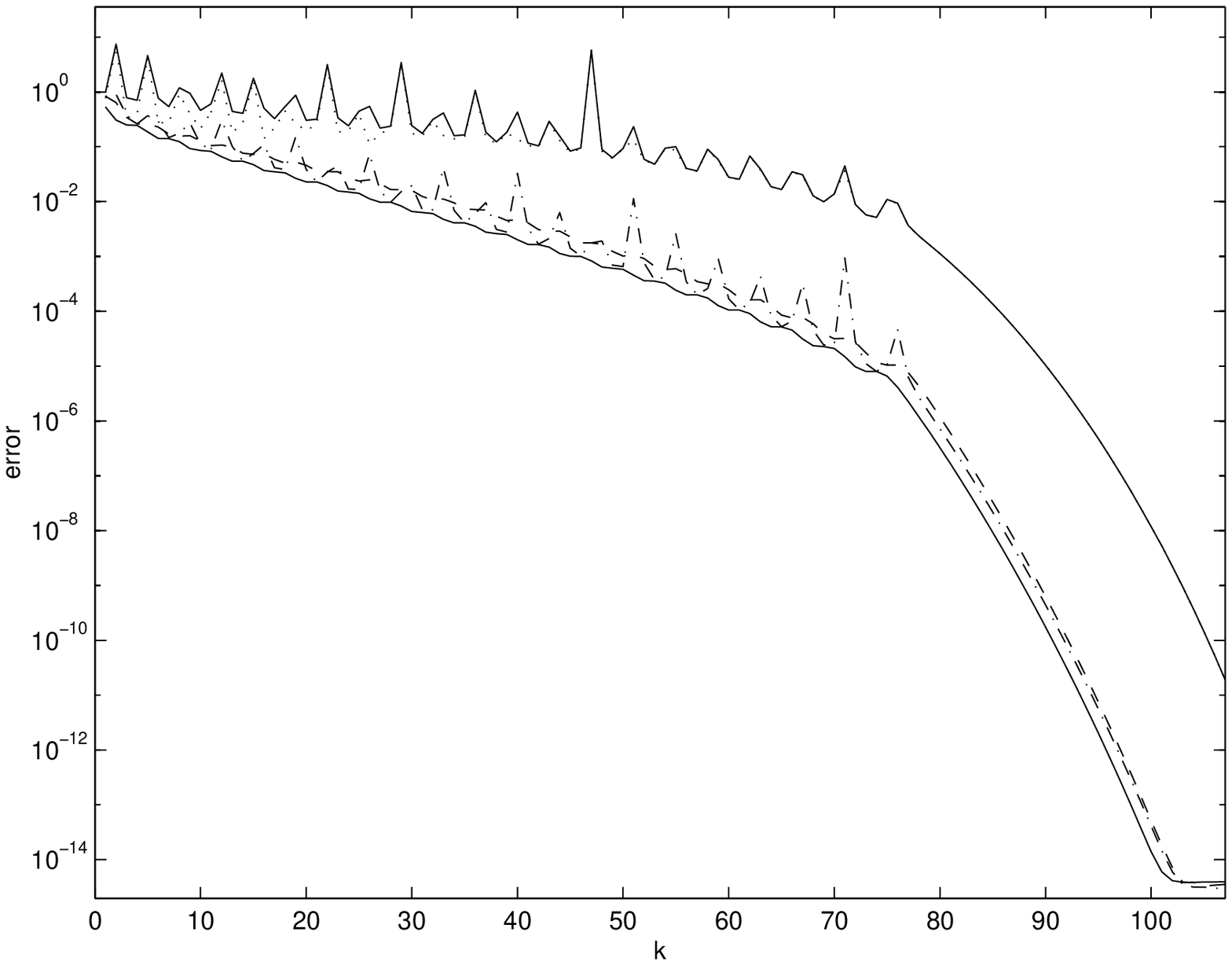}
\caption{Above: Lanczos without extra orthogonalization, below:
  Lanczos with extra orthogonalization.
  optimal (solid lower), Eq. (\ref{borici1eq}) (dotted),
  Eq. (\ref{vdveq}) (dash dot),
Eq. (\ref{harsigneq}) (dashed), norm of the CG residual (solid top)}
\label{lanczosfig}
\end{figure}

Returning to the context of the overlap
operator with Lanczos relation (\ref{lanczosmfQ}) the harmonic
Lanczos approximation becomes
\begin{equation}
\label{harsigneq}
\hbox{sign}(Q)b \approx V_k \hbox{sign}(T_k + z e_k^{\dagger}) e_1 \|b\|_2 \hbox{ with }
z = \beta_k^2 T_k^{-1}e_k.
\end{equation}
We now have three methods to approximate $\hbox{sign}(Q)b$ based on Lanczos for $Q$.
We demonstrate their differences for a simple diagonal matrix $Q$. The diagonal of $Q$
contains the elements $-30$,$-29$, $\dots$,$-10$, $1$,$2$,$\dots$,$100$. The vector $b$
has unit-length and all its components are equal.
In Figure \ref{lanczosfig} we plotted
the error as a function of $k$. The error of the optimal method is
the error of the `best you can do method' from Section \ref{optsec}, computed as 
$\|(I-V_{k+1} (V_{k+1}^{\dagger}V_{k+1})^{-1} V_{k+1}^{\dagger})\hbox{sign}(Q)b\|$.

From Figure \ref{lanczosfig} a few features are apparent. In the
first place, the loss of orthogonality between the $v_i$'s in the
Lanczos algorithm due to finite precision arithmetic only delays
the convergence, see also \cite{DGK98}. The error for (\ref{borici1eq}) coincides after
a while (in norm) with the residual for the CG process. The
harmonic Lanczos approximation indeed shows a smoother
convergence than the Lanczos approximation but seems in general
slightly less accurate. Finally, we note that (\ref{vdveq}) is in
every other step almost optimal and superior to the other
approaches.
Both (\ref{vdveq}) and (\ref{borici1eq}) implicitly construct polynomials
that interpolate the sign-function on the Ritz values. 
The approximation in (\ref{borici1eq}) uses the additional 
degree of freedom for a root in zero. 
A possible explanation for the better results of
(\ref{vdveq}) in comparison with (\ref{borici1eq}) is that 
this additional root is a restriction.
Further analysis is necessary to understand these phenomena.

\subsection{The quality of the polynomials} \label{qualsec}

\begin{figure}
\centering
\includegraphics[width=\textwidth]{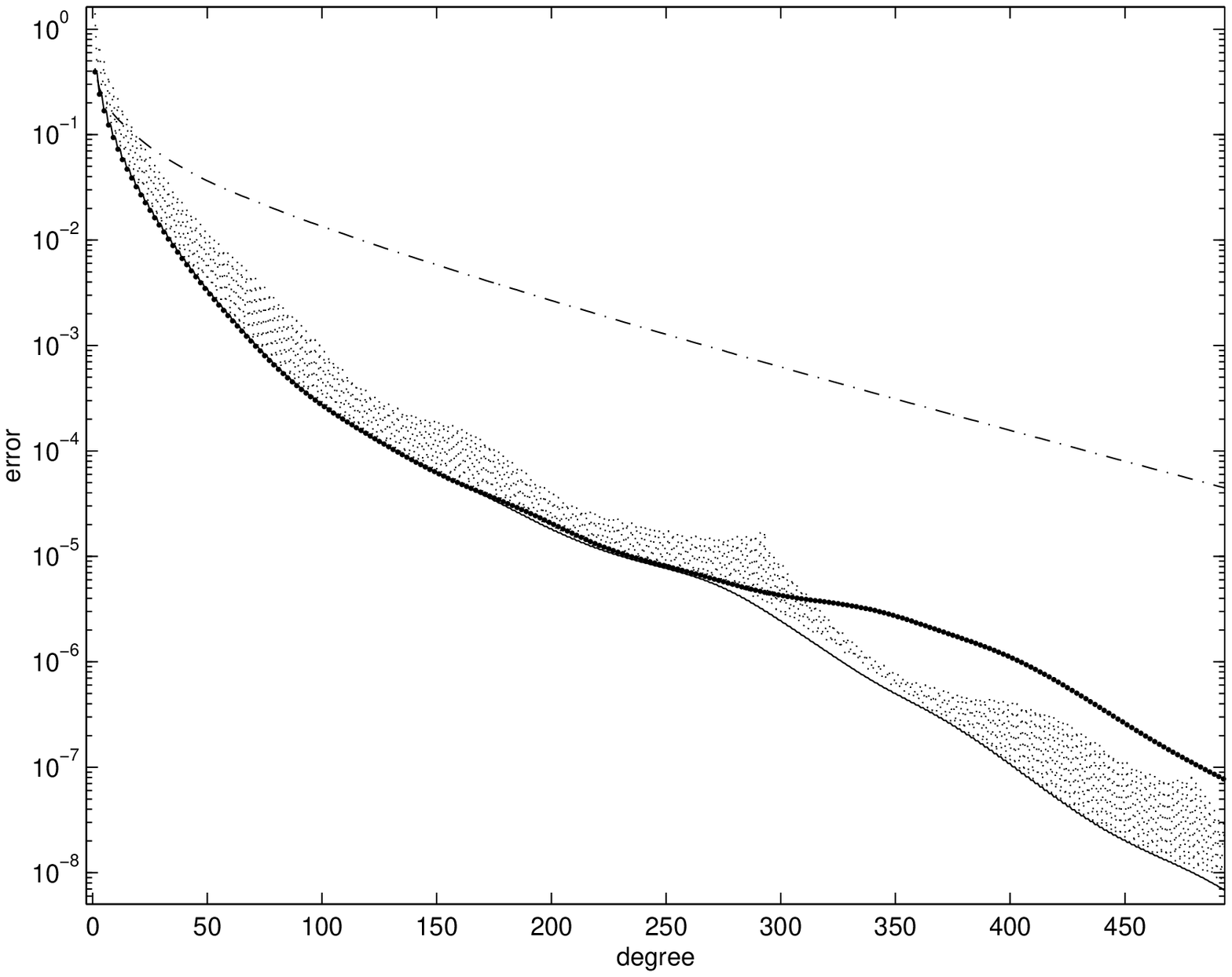}
\includegraphics[width=\textwidth]{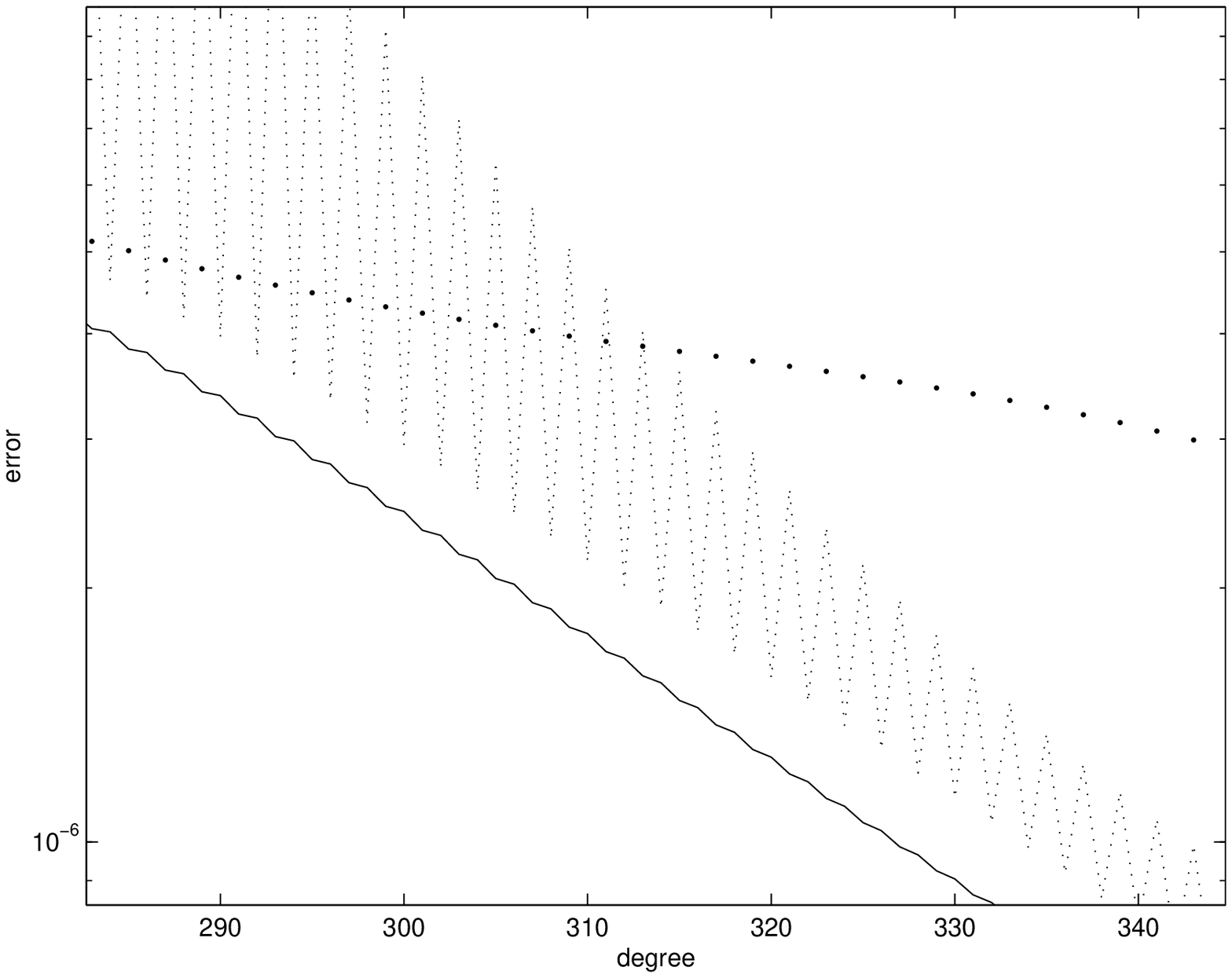}
\caption{The error as a function of the polynomial degree: optimal (solid),
 Chebyshev method (-.), Eq. (4.7) (dots), Eq. (4.5) (dotted). The second 
diagram shows a detail of the first one. 
}
\label{qualfig}
\end{figure}

In this section we compare several ways to construct polynomial
approximations to the $\hbox{sign}$-function by methods from the
different classes that we discussed. The methods are the `best' method
from Section \ref{optsec}, the Chebyshev approach from Section
\ref{statsec}, Lanczos on $Q^2$ with (\ref{borici2eq}) and
Lanczos on $Q$ with (\ref{vdveq}). The results for a realistic
problem in QCD are given in Figure \ref{qualfig}. The lattice is
$16^4$, $\kappa = 0.208$ and $\beta = 6$.

All methods considered are not too far away from the `best you can do'
method.  It thus appears that for realistic QCD matrices, $Q$, it is
not a real drawback to restrict the approximation to the class of odd
polynomials (i.e.\ to the space $QK_{k/2}(Q^2,b)$), as might have been
anticipated from the discussion in Section \ref{optsec}.  Therefore, 
we consider in the remainder of this paper `Lanczos on $Q^2$' with 
(\ref{borici2eq}) as an efficient method within the class of Lanczos
approximations.

\subsection{Error estimation}

An important question is when to terminate Alg. \ref{lanczosalg}
such that the approximation (\ref{borici2eq}) is within a distance
$\epsilon$ to the exact vector $\sign{Q}\cdot b$. In \cite{Bor99c} it is proposed
to use the norm of the residual of the related CG process as an
upper bound for the error in $2$-norm. The norm of this residual
can be computed at little additional cost in Alg.
\ref{lanczosalg}, see \cite{Bor99c} for details. Note that no theoretical
justification for this stopping criterion was given in \cite{Bor99c}.
In this section
we will {\em prove} that the norm of the CG residual is always larger than
the error for (\ref{borici2eq}). This result is formulated in
Theorem \ref{boricibound}.

For the proof we will use the following integral representation
of the inverse square root, see, e.g., \cite{Bor99a}:
\begin{equation}
\label{integraleq}
A^{-1/2} = \frac{2}{\pi}\int_{t=0}^{\infty} (A+t^2 I)^{-1} \dt .
\end{equation}
Furthermore, we will exploit the following convenient property of
the conjugate gradient method applied to a shifted system.
\begin{lem}
\label{shiftlemma}
Let $r_k$ denote the CG residual
for $k$ steps CG when solving $Ax=b$
and, similarly, $r_k^{\tau}$ for solving $(A+\tau I)x=b$, both methods starting
with the initial iterate zero. Then
$$
r_k^{\tau} = \phi_k^{\tau} r_k
$$
with
$$
\phi_k^{\tau} \equiv \prod_{j=1}^k \frac{\theta_{j,k}}{\theta_{j,k}+\tau},
$$
where $\theta_{j,k}$ denotes the $j$-th eigenvalue of $T_k$.
\end{lem}
\begin{pf}
From e.g. \cite{PPV95} we have the following polynomial characterizations
for the residuals
$$
\begin{array}{ll}
r_k = \pi_k(A)b/\pi_k(0), & \pi_k(t) \equiv \det(t I - T_k) \\
r_k^\tau = \pi_k^\tau(A+\tau I)b/\pi_k^{\tau}(0), & \pi_k^\tau(t) \equiv
\det(t I - (T_k+\tau I)  ) .\\
\end{array}
$$
Hence, $$r_k^{\tau} = \pi_k^\tau(A+\tau I)b/\pi_k^{\tau}(0) = \phi_k^\tau 
\pi_k(A)b/\pi_k(0) = \phi_k^{\tau}r_k.$$
\end{pf}

We are now in a position to formulate the main theorem in this
section.
\begin{thm}
\label{boricibound}
Let $\det(Q)\ne 0$. Then
\begin{equation}
\label{result2}
\| \hbox{\rm sign}(Q)b - Q\hV_k \hT_k^{-1/2} e_1 \|_2
\le
\|r_k\|_2
\le
2
  \kappa \left( \frac{\kappa-1}{\kappa+1} \right)^k \cdot \|b \|_2 ,
\end{equation}
where $\kappa \equiv \|Q\|_2 \|Q^{-1}\|_2$ and $r_k$ is the
residual in the $k$-th step of the CG method applied to the system $Q^2x = b$
(with initial residual $b$, i.e.\ initial guess $0$).
\end{thm}
\begin{pf} The expression $\hbox{\rm sign}(Q)b - Q\hV_k \hT_k^{-1/2} e_1$
can be rewritten with (\ref{integraleq}) as
$$
\frac{2}{\pi}\int_0^{\infty}Q (t^2 I+Q^2)^{-1}b -
Q\hV_k(t^2 I+\hT_k)^{-1}\hV_k^{\dagger}b \ \dt =
$$
$$
\frac{2}{\pi}\int_0^{\infty}Q(t^2 I+Q^2)^{-1} (t^2 I+Q^2)
\left((t^2 I+Q^2)^{-1} - \hV_k(t^2 I+\hT_k)^{-1}\hV_k^{\dagger}\right)b \
\dt =
$$
$$
\frac{2}{\pi}\int_0^{\infty}Q(t^2 I+Q^2)^{-1} r^{t^2}_k \
\dt ,
$$
where $r^{t^2}_k \equiv b - (Q^2+t^2 I)\hV_k (\hT_k+t^2
I)^{-1}e_1$. With Lemma \ref{shiftlemma} this can be expressed as
\begin{equation}
\label{finalboundeq} \frac{2}{\pi}\int_0^{\infty}Q(t^2
I+Q^2)^{-1} \phi_k^{t^2} r_k \ \dt = X r_k \ \hbox{, with}\ X \equiv
\frac{2}{\pi}\int_0^{\infty}Q(t^2 I+Q^2)^{-1} \phi_k^{t^2} \
\dt
.
\end{equation}
Then the first bound in (\ref{result2}) follows by bounding the
eigenvalues of the operator $X$ in (\ref{finalboundeq}). To this purpose,
note that the eigenvalues of $X$ are given as
\begin{equation} \label{evintegraleq}
\frac{2}{\pi}\int_0^{\infty}\lambda_i(t^2+\lambda_i^2)^{-1} \phi_k^{t^2} \ \dt
\end{equation}
with $\lambda_i$ the eigenvalues of $Q$.  For $\lambda_i$ fixed, the
integrand in (\ref{evintegraleq}) has constant sign. Since $0\le
\phi_k^{t^2}\le 1$ we therefore get
\[
\left| 
\frac{2}{\pi}\int_0^{\infty}\lambda_i(t^2+\lambda_i^2)^{-1} \phi_k^{t^2} \ \dt
\right| \; \leq \; \left| \frac{2}{\pi}\int_0^{\infty}\lambda_i
(t^2+\lambda_i^2)^{-1} \ \dt \right| \; = \; 1.
\]
This proves $ \|X\|_2 \leq 1$ and thus $\|Xr_k\|_2 \leq \|r_k\|_2$. 

The {\em a priori} upper bound follows directly from {\em a
priori} error estimates for CG, as in Theorem 3.1.1 in
\cite{Gre97a}.
\end{pf}

The question is how tight the lower bound in (\ref{result2}) may
be. After one step of Alg.~\ref{lanczosalg}, i.e.\ for $k=1$, where we
have a simple expression for $\phi_1^{t^2}$, we can explicitly
integrate (\ref{evintegraleq}). As a result we then see that the
residual is overestimated by at most a factor
$1+\frac{\lambda}{\sqrt{T_1}}$, in the direction of the eigenvector
corresponding to $\lambda$.  Hence, we are at most a factor
$1+\kappa(Q)$ off and at least a factor $1+\kappa(Q)^{-1}$.

When a precision $\epsilon$ is required, the process can be safely
terminated as soon as $\|r_k\|_2\le\epsilon$.

\subsection{Practical implementations} \label{pracsec}

In practice it is not feasible to store all $k$ vectors $\hv_i$ for
the evaluation of (\ref{borici2eq}). This problem can be circumvented
by executing Alg.~\ref{lanczosalg} twice \cite{Bor99a,Bor99c}. In the
first step the tridiagonal matrix $\hT_k$ is constructed after which
the vector $\hT_k^{-1/2}e_1\|b\|_2$ is computed. In the second run of
Alg.  \ref{lanczosalg} the $\hv_i$'s are combined to compute $\hV_k
\hT_k^{-1/2}e_1\|b\|_2$. This doubles the number of MVs but, as
remarked by Neuberger \cite{Neu00}, this method can be still
competitive because of a potentially better exploitation of the
computer memory hierarchy (e.g., cache effects).

Another issue is the computation of $\hT_k^{-1/2}e_1\|b\|_2$. Our own
numerical experiments show that a computation with a full spectral
decomposition, as in \cite[Algorithm 2]{Bor99c}, is too CPU-time
consuming. Instead it seems natural to exploit the special tridiagonal
structure of $\hT_k$ in a similar way as done for the exponential
function in \cite{Sa92}, for example. In our numerical experiments we
computed $\hT_k^{-1/2}e_1\|b\|_2$ using a rational approximation
expanded as the sum over poles (called a {\em partial fraction
expansion}, {\em PFE})
$$
\hT_k^{-1/2} \approx \sum_{i=1}^m \omega_i (\hT_k+\tau_i I)^{-1}.
$$
A discussion of the choice of suitable coefficients $\tau_i$ and
$\omega_i$ can be found in Section \ref{ratsec}.  We are now required
to solve $m$ tridiagonal linear systems $(\hT_k+\tau_i I) z = e_1
\|b\|_2$ and this can be done efficiently with the {\em LAPACK}
function {\em DPTSV}. This results in $O(km)$ flops which can be much
less than the $O(k^2)$ flops that are needed for the computation of a
full spectral decomposition. In the remainder of this paper we will
refer to this implementation of (\ref{borici2eq}) as Lanczos/PFE.

\section{The PFE/CG Method} \label{pfesec}

The Lanczos approximations in the previous section require two passes of the Lanczos method
to limit memory requirements. This doubles the number of matrix-vector multiplications.
An elegant idea for working with only a fixed number of vectors without requiring two
passes was proposed,
in the context of QCD, by Neuberger \cite{Neu00}. Suppose $r(t)$ is a rational
function approximating the $\hbox{sign}$-function in the interval $[-b,-a]\cup[a,b]$ that
contains the eigenvalues of $Q$. Let
$r(t)$  be represented by the following partial fraction expansion
\begin{equation}
\label{pfesigneq}
\sign{t}\approx r(t) =\sum_{i=1}^m \omega_i \frac{t}{t^2+\tau_i},
\end{equation}
then
\begin{equation}
\label{pfeeq}
\sign{Q}b  \approx x^{\mbox{pfe}} \equiv r(Q)b = \sum_{i=1}^m \omega_i \frac{Q}{Q^2+\tau_i I}b.
\end{equation}
The idea is now to solve the $m$ linear systems in (\ref{pfeeq}) with a 
{\em multi-shift Conjugate Gradient} method \cite{FM99,Je96}. 
The philosophy behind the multi-shift CG method is the following Lanczos relation
for shifted matrices
\begin{equation}
\label{lancrelshift}
(Q^2+\tau_i I)\hV_k =  \hV_k(\hT_k+\tau_i I) + \hbeta_k \hv_{k+1} e_k^{\dagger}.
\end{equation}
Hence, the orthonormal basis has to be constructed only once 
for the various shifts and after $k$ steps Lanczos we can construct
$m$ approximate solutions 
and the corresponding residuals as
$$
x_k^i \equiv \hV_k (\hT_k + \tau_i I)^{-1} e_1 \|b\|_2\hbox{ and }
r_k^i \equiv b - (Q^2+\tau_i I)x_k^i.
$$ 
Similar in spirit to other Krylov subspace methods for shifted linear systems
(\cite{FM99,Fretal95,GGT96,Je96}, e.g.) the multi-shift CG method computes these 
vectors in an efficient way. 
The final approximation to $\sign{Q}b$ now reads
\begin{equation}
\label{finaleq}
\sign{Q}b \approx x^{\mbox{pfe}} \approx x_k \equiv 
  \sum_{i=1}^m \omega_i Q x_k^i.
\end{equation}
We refer to this method as PFE/CG 
(of course, in a practical implementation the solutions of the $m$ different systems are  
not stored but are immediately combined to save memory).
We are again in the Krylov subspace framework described in 
Section~\ref{optsec}, since $x^{\mbox{pfe}}$ is from the Krylov subspace
$K_l(Q,b)$ with $l = 2k+1$, $k$ being the maximum number of iterations
in multi-shift CG. Note also that this method is, for the same shifts and poles,
mathematically equivalent to the Lanczos/PFE method from Section \ref{pracsec}.

\subsection{Error estimation}
\label{apostsec}
We need a criterion for the termination of the multi-shift CG method such that
$x_k$ in (\ref{finaleq}) is close enough to $\sign{Q}b$, or 
$\|\hbox{sign}(Q)b-x_k\|\le \epsilon$.
The error in the PFE/CG method consists of two parts. 
First, we demand  the following accuracy of the rational function (this can be
cheaply monitored in its construction)
\begin{equation}
\label{pfeacc}
|\sign{t} - r(t) |
\le \epsilon/2 \hbox{ for } t \in \spec(Q)
\subset [-b,-a] \cup [a,b].
\end{equation}
This gives
$$
\|\sign{Q}b - x^{\mbox{pfe}} \|_2 \le \epsilon/2 .
$$
Furthermore, we require that
$$
\|x^{\mbox{pfe}} - x_k \|_2 \le \epsilon/2.
$$
Our termination condition consists of checking if this condition is fulfilled. 
We use the following theorem for this which is similar in spirit to Theorem \ref{boricibound}.
\begin{thm}
\label{apostpfecg}
Let the rational approximation satisfy (\ref{pfeacc}) and 
$\omega_i\ge 0$ for $i\in \{1,\dots,m\}$ and 
$0 < \tau_1 \le \tau_2 \le \dots \le \tau_m$,
then 
$$
\| x^{\mbox{\rm pfe}} - x_k\|_2 \le (1+\epsilon/2) \|r_k^1\|_2 .
$$
\end{thm}
\begin{pf}
$$
 x^{\mbox{pfe}} - x_k  =
 \sum_{i=1}^m \omega_i Q \left( \frac{I}{Q^2+\tau_i I}b - x_k^i \right)   =
\sum_{i=1}^m \omega_i \frac{Q}{Q^2+\tau_i I} r_k^i   
$$
$$
= \sum_{i=1}^m \omega_i \frac{Q}{Q^2+\tau_i I} \phi_k^{(\tau_i-\tau_1)}
r_k^1  = Xr_k^1 \hbox{ with } 
X \equiv  \sum_{i=1}^m \omega_i \frac{Q}{Q^2+\tau_i I} \phi_k^{(\tau_i-\tau_1)}
$$
Here, we have used Lemma \ref{shiftlemma}.
The proof follows by noting that the eigenvalues of $X$ are 
of the form $\sum_{i=1}^m \omega_i \frac{t}{t^2 + \tau_i} \phi_k^{(\tau_i
-\tau_1)}, \; t \in \spec(Q)$ and this expression is bounded in absolute value
by $|r(t)|$. 
From (\ref{pfeacc}) we 
find  that $|r(t)| \le 1+\epsilon/2$ for 
$t \in \spec(Q)$.
\end{pf}
So, we can terminate the multi-shift CG method when the residual of the first system
satisfies
\begin{equation}
\|r_k^1\|_2 \le
\frac{\epsilon}{2+\epsilon}.
\label{eps2}
\end{equation}
Then the error of $x_k$ is bounded as  $\|x_k -\hbox{sign}(Q)b\|_2 \le \epsilon$.

\subsection{The choice of the rational approximation} \label{ratsec}

For the PFE/CG method the cost of computing $\sign{Q}b$ is basically the cost
of one run of CG plus some additional cost for updating of the $m-1$ additional
systems. This puts emphasis
on the efficiency of the used rational approximation. Let us first make our
terminology precise: We consider a given function $f$ which is defined
on a set $D$ and we assume that we have a space $S$ of approximating functions, all defined
on $D$. Then we call $g \in S$ a {\em best approximation} for $f$ on $D$ from $S$,
if $g$ minimizes the quantity
\[
\sup_{t \in D} |f(t) - h(t)|
\]
among all functions $h$ from $S$. In our context, $S$ will be a space
$R_{i,j}$ of rational functions $r(t) = p(t)/q(t)$ with $p \in \Pi_i$ and 
$q\in\Pi_j$.

In his original proposition, Neuberger \cite{Neu00} uses the following rational
approximation from $R_{2m-1,2m}$
$$
r(t) = \frac{(t+1)^{2m}-(t-1)^{2m}}{(t+1)^{2m}+(t-1)^{2m}},
$$
which can be written in the form of (\ref{pfesigneq}) with
$$
\omega_i = \frac{1}{m} \cos^{-2}\left(\frac{\pi}{2m}(i-\frac{1}{2})\right), \quad
\tau_i = \tan^2\left(\frac{\pi}{2m}(i-\frac{1}{2})\right).
$$
It can be easily checked that this approximation is exact for $|t|=1$ and 
that the error for $|t|\ge 1$ is increasing for increasing $|t|$.
From this and $r(t)=r(t^{-1})$ it follows that this rational function  
approximates $\hbox{sign}(t)$ well on sets $D$ of the form $[-1/c,-c]\cup[c,1/c]$
for some specific value $c\in (0,1]$ (independent of $m$).
Therefore, it is common practice
to map the interval $[-b,-a]\cup[a,b]$ to  a range of the form
$[-1/c,-c]\cup[c,1/c]$ by scaling
with a factor $(ab)^{-1/2}$, yielding $c=\sqrt{b/a}$. 

Another consequence is that the error is maximal
for $t=\sqrt{\kappa}$ (see also \cite{EHN98}).
Using this it follows that for a precision of $\epsilon/2$ 
in this scaled interval we need $m$ poles where $m$ is some integer with
\begin{equation}
\label{mboundeq}
m \ge \frac{1}{2}
\log\left(\frac{\epsilon}{4-\epsilon}\right) \left/ \log\left(\frac{\sqrt{b/a}-1}{\sqrt{b/a}+1}\right)    \right.  .
\end{equation}
From (\ref{mboundeq}) it appears that the number of required poles can
be quite large. The function $r(t)$ is {\em not} a best approximation
in the sense defined before, see Proposition~\ref{Zolprop} below.

A different idea is to construct an approximation of the form
$\sign{t}\approx t s(t^2)$ where
\begin{equation}
\label{invsqrtpfe}
t^{-1/2} \approx s(t) = \sum_{i=1}^m \omega_i \frac{1}{t+\tau_i} \hbox{ for } t\in[a^2,b^2].
\end{equation}
In \cite{EHN98}, Edwards et al.\ propose to construct a best
approximation for $t^{-1/2}$ on $[a^2,b^2]$ from $R_{m,m}$ by means of
the Remez method and to compute the $\omega_i$ and $\tau_i$ from this
expression (an additional constant term can be necessary).  In the
following we will refer to this methods as EHN-approach
(Edwards-Heller-Narayanan). Note that whilst $s(t)$ {\em is} a best approximation
to the {\em inverse square root}, $t\cdot s(t^2)$ {\em is not} a best
approximation to the {\em $\hbox{sign}$-function} on
$[-b,-a] \cup [a,b]$. 

We now propose to use a rational approximation to the sign-function which is the
best approximation on $D = [-b,-a] \cup [a,b]$. An explicit representation 
of this best approximation is due to Zolotarev. His work 
was brought to our attention by the paper of Ingerman et al.
\cite{IDK00} and seems not yet have been applied in the context of the
overlap operator.
The key point is that finding the optimal approximation from $R_{2m-1,2m}$ to
the $\hbox{sign}$-function on $[-b,-a] \cup [a,b]$, is equivalent to
finding the best rational approximation in {\em relative sense} from $R_{m-1,m}$
to the inverse square root on $[1,(b/a)^2]$. This is expressed by the following
proposition.
\begin{prop}[Zolotarev \cite{PPo87}] \label{Zolprop}
Let $s \in R_{m-1,m}$ be the best relative approximation to $t^{-1/2}$ on the set $[1,(b/a)^2]$, i.e. the function which minimizes
\[
\max_{t \in [1,(b/a)^2]} | \left( t^{-1/2} - f(t) \right) / t^{-1/2} |
\; = \;  \max_{t \in [1,(b/a)^2]}
   \left| 1 - \sqrt{t}\cdot f(t)\right|  
\]
over all $f \in R_{m-1,m}$.
Then the best approximation to the $\hbox{sign}$-function on $[-b/a,-1] \cup [1,b/a]$
from $R_{2m-1,2m}$ is given by
$$
r(t) = ts(t^2).
$$
and, consequently, the best approximation to the \hbox{sign}-function on $[-b,-a] \cup [a,b]$
from $R_{2m-1,2m}$ is $r(at)$.
\end{prop}

Zolotarev furthermore showed that this rational approximation $s(t)$ is
explicitly known in terms of the Jacobian elliptic function $\hbox{sn}$,
so there is no need for running the Remez algorithm. Moreover, the
number of poles required for a given accuracy will turn out
to be significantly smaller than for the previous two discussed approaches.

\begin{thm}[Zolotarev \cite{IDK00,PPo87}]
\label{zolth}
The best relative approximation $s(t)$ from $R_{m-1,m}$ for $t^{1/2}$
on the interval $[1,(b/a)^2]$ is given by
\begin{equation} \label{sdefeq}
s(t) = D \frac{\prod_{i=1}^{m-1}(t+c_{2i}) }{\prod_{i=1}^m(t+c_{2i-1}) },
\end{equation}
where
$$c_i = \frac{\hbox{\rm sn}^2\left(iK/(2m); \sqrt{1-(b/a)^2}\right)}
{1-\hbox{\rm sn}^2\left(iK/(2m); \sqrt{1-(b/a)^2}\right)},$$
$K$ is the complete elliptic integral and
$D$ is uniquely determined by the condition
$$
\max_{t\in [1,(b/a)^2]} \left( 1 - \sqrt{t} s(t)\right)  = -\min_{t\in [1,(b/a)^2]}
\left( 1 - \sqrt{t} s(t) \right).
$$
\end{thm}

From the above 
theorem we can derive the coefficients for (\ref{invsqrtpfe}) and
(\ref{pfesigneq}).  Its use can drastically reduce the number of
required poles to achieve a certain accuracy, compared with the two
other rational approximations discussed before.  For the three
rational approximations Table \ref{tabel1} gives the number of
poles necessary  
to achieve an accuracy of $0.01$.
Unfortunately, as far as we know, for the EHN and Zolotarev
approximations there is no real a priori knowledge of the required
number of poles for a given accuracy. The number of
required poles in Table \ref{tabel1} for the EHN-approximation are
 taken from \cite{EHN98} and for the Zolotarev method we have used a
 simple numerical technique.
\begin{table}
\begin{center}
\caption{Number of poles necessary to achieve accuracy of $0.01$
\label{tabel1}}
\begin{tabular}{l|llll}
$b/a$ & Neuberger & EHN & Zolotarev \\
\hline
$200$    & $19$ & $7$   & $5$ \\
$1000$   & $42$ & $12$  & $6$ \\
\end{tabular}
\end{center}
\end{table}

\subsection{Removing converged systems} \label{remsec}

In the preceding section we tried to reduce the number of poles, $m$,
by choosing a high quality rational approximation for the
$\hbox{sign}$-function.  A supplementary idea results from a closer
look at the shifts $\tau_i$.  It appears that some of them are quite
large and from Lemma \ref{shiftlemma} we see that the corresponding
residuals in the multi-shift CG method become very small quickly.  As
in the proof of Theorem \ref{apostpfecg}, we can write the error in
step $k$ as
$$
 x^{\mbox{pfe}} - x_k  =
 \sum_{i=1}^m \omega_i Q \left( \frac{I}{Q^2+\tau_i I}b - x_k^i \right)   =
\sum_{i=1}^m \omega_i \frac{Q}{Q^2+\tau_i I} r_k^i .
$$
This shows that systems with a sufficiently small residual contribute
little to the error and apparently they need not be solved as
accurately.  It is obvious that, if we want an accuracy of
$\epsilon/2$ as in Section \ref{apostsec}, we can start neglecting
system $j$ after iteration $k$ when
$$
\|\omega_j \frac{Q}{Q^2+\tau_j I} r_k^j\| \le g_j \frac{\epsilon}{2},
$$
where $g_i > 0$ and $\sum_{i=1}^m g_i = 1$.  By using that $\|\omega_j
Q (Q^2+\tau_j I)^{-1}\| \le \left| \omega_j \right| / (2\sqrt{\tau_j})$, we find that
we can stop updating system $j$ if
\begin{equation}
\|r_k^j\|_2 \le \epsilon g_j \frac{\sqrt{\tau_j}}{\omega_j}.
\label{eps3}
\end{equation}
We note that this idea can be seen as an alternative for the
termination condition in Section \ref{apostsec}. In general it will
require more CG iterations but with less updates for the additional
systems.  For numerical results we refer to Section \ref{numexpsec}.
In our implementation we have taken $g_i = 1/m$, but we remark that
more sophisticated choices are possible, for example $\tau$-dependent
ones.

\section{Numerical Experiments} \label{numexpsec}

In this section we report on the performance of some of the discussed
methods for realistic configurations in QCD. All the experiments are
carried out on the cluster computer AliCE installed at Wuppertal
University \cite{PIK:2002}.  

We work with quenched configurations of size $16^4$ which results in a
matrix $Q$ with $786 432$ complex valued unknowns.  
The value of $\kappa$ has been
chosen as $0.208$. This corresponds to a mass parameter $m=-1.6$ for the
Wilson-Dirac argument of $D$, a standard choice at inverse coupling
$\beta=6.0$.  For our experiments we have taken $5$ statistically
independent configurations. The first two columns of Table
\ref{confstab} give the smallest and the largest eigenvalue of $Q$ (in
modulus) as $a$ and $b$, respectively. These numbers were used for the
defining intervals of the rational approximations.
\begin{table}
\begin{center}
\caption{The spectral properties of the configurations used and the
required number of poles for a precision of $10^{-10}$.\label{confstab}}
\begin{tabular}{c|cccc}
Conf. & $a$ & $b$  & Poles
Neub. & Poles Zol.\\ \hline
1 &4.548(-3) & 2.4819  &143 & 21\\
2 &1.385(-2)  & 2.4818  &82 & 18\\
3 &1.169(-2)  & 2.4825  &89& 19\\
4 &2.226(-2)  & 2.4824  &65& 17\\
5 &3.024(-2)  & 2.4819  &56 & 16\\
\end{tabular}
\end{center}
\end{table}      

We have computed $\sign{Q}b$ with a precision $\epsilon$ of at least
$10^{-10}$, see \eqs{result2}{eps2}{eps3}. We compare $5$ different
approaches: The Chebyshev approach from Section \ref{statsec}, the
(two pass) PFE/Lanczos method as described in Section \ref{pracsec}
with the Zolotarev coefficients (as derived from Theorem \ref{zolth}),
the standard PFE/CG method with the termination condition from Section
\ref{apostsec} with the coefficients used by Neuberger and with the
Zolotarev coefficients, and the PFE/CG method with the stopping idea
from Section \ref{remsec} with the Zolotarev coefficients.  The number
of poles required for the two partial fraction expansions is reported
in Table \ref{confstab}.  All benchmark results are summarized in
Table \ref{bench1}. The number of processors used is $16$.
\begin{table}[!htb]
\begin{center}
\caption{Benchmarks.\label{bench1}}
\begin{tabular}{l|lllll}
Conf. & 1 & 2 & 3 & 4 & 5 \\ \hline 
\multicolumn{6}{c}{Chebyshev}\\
\hline
MVs & 9501 & 3501 & 4001 & 2301 & 2201 \\
time/s & 655 & 247 & 278 & 160 & 154 \\
\multicolumn{6}{c}{Lanczos/PFE}\\
\hline
MVs & 2281 & 1969 & 1953 & 1853 & 1769 \\
time/s & 150 & 131 & 129 & 124 & 118 \\
\multicolumn{6}{c}{PFE/CG Neuberger}\\
\hline
MVs & $\times$ & 985 & 977 & 929 & 887 \\
time/s & $\times$ & 340 & 362 & 274 & 215 \\
\multicolumn{6}{c}{PFE/CG Zolotarev without removal}\\
\hline
MVs        & 1141 & 985  & 977 & 927 & 885 \\
 time/s & 154  & 125  & 125 & 116 & 102 \\
\multicolumn{6}{c}{PFE/CG Zolotarev with removal}\\
\hline
MVs & 1205 & 1033 & 1033 & 971 & 927 \\
time/s & 122 & 93 & 97 & 87 & 79 \\
\end{tabular}
\end{center}
\end{table}

For the first configuration (where $a$ is small), we were not able to
run the PFE/CG method with Neuberger's approximation. This is due to
the fact that we had too many poles in the PFE, so memory requirements
became too large. (Besides the memory requirement for CG we would have had 
to store $142$ additional vectors).

From Table \ref{bench1} we see that Chebyshev needs the largest number
of matrix-vector multiplications. However, the additional work per
iteration is quite small in Chebyshev, so the execution time of
Chebyshev is smaller than that of the PFE/CG Neuberger method.
Chebyshev turns out to be most sensitive to the ratio $b/a$: while the
iteration numbers of all other methods depend only quite moderately on
$b/a$, Chebyshev requires an iteration number which is approximately
proportional to $b/a$.

The Lanczos/PFE method needs about 25\% less matrix-vector
multiplications than Chebyshev on configurations 4 and 5 and substantially
less on configurations 1, 2 and 3. Note that our results for
Lanczos/PFE are given for the two-pass methods. If we could store all
Lanczos vectors, the number of matrix-vector multiplications would
decrease by a factor of 2!

The partial fraction expansion methods all require a similar number of
matrix vector multiplications. The computational overhead in the
shifted CG method depends linearly on the number of poles in the
PFE. This is the reason why the (optimal) Zolotarev rational
approximation results in a much lower execution time than Neuberger's
rational approximation. Thus, Zolotarev saves execution time as well as
computer memory. Without the early removal of converged systems,
Zolotarev requires an execution time similar to the (two pass)
Lanczos/PFE method. However, with this removal, Zolotarev saves
another 20\% to 25\% in execution time and thus turns out to be the
overall best of all methods considered.

In practical QCD experiments using the Chebyshev method, it has been
suggested to speed up Chebyshev convergence by first computing some
low eigenvectors and then projecting the configuration onto the
orthogonal complement. In this manner, the value for $a$ to be used in
Chebyshev can be increased substantially.

In this context, the following comparison `across the configurations'
between Chebyshev and Zolotarev with early removal is particularly
noteworthy: The execution time of Chebyshev on Configuration 4 or 5 is
still more than 30\% higher than the time for Zolotarev on
Configuration 1 (and all other configurations).  Looking at the
corresponding values for $a$, we see that even if for Configuration 1
we were able to `project out' all eigenvalues and eigenvectors in the
range from $4.548 \cdot 10^{-3}$ (the smallest eigenvalue) up to $3
\cdot 10^{-2}$, we would still not obtain a better performance for
Chebyshev (using the projected system) as compared to Zolotarev.

\section{Summary and Outlook}

We have improved upon known methods and have presented novel ideas to
compute the sign-function of the hermitian Wilson matrix within
Neuberger's overlap fermion prescription.  Our comparisons on
realistic quenched gauge configurations demonstrate that the PFE/CG
method with removal of converged systems, which is based on
Zolotarev's theorem, is superior to other PFE/CG procedures so far
applied in the literature.  The Zolotarev approach is the provably
best rational approximation to the sign function on domains of the form
$[-b,-a] \cup [a,b]$ and therefore requires the smallest number of poles.
As this rational
approximation is explicitly known in terms of elliptic functions,
we can avoid to run Remez' algorithm.  

As a major result of this work, we have derived explicit error bounds
for both Lanczos and PFE methods that allow for safe termination of
the respective iterative processes. This is a mandatory requirement
for a controlled two-level iteration in the overlap scheme. 
In future work we will concentrate on improving the coupling
between the progress of the outer and the accuracy of
the inner iteration and we are going to include the effects of projecting
out low-lying eigenmodes of the hermitian Wilson-Dirac matrix $Q$.

\section*{Acknowledgments}
J.\ v.\ d.\ E.\ was partly supported by the EU Research and Training
Network HPRN-CT-2000-00145 ``Hadron Properties from Lattice QCD''. We
thank N.\ Eicker for sharing his QCD library and his help with the
Wuppertal cluster system ALiCE.

\begin{appendix}
\section{Definitions}
The hopping term of the Wilson-Dirac matrix reads:
\begin{eqnarray}
 \label{HOPPING}
  {D_W}_{nm} = \frac{1}{2}\sum_{\mu=1}^4
    (I-{
    \gamma}_{\mu}){U}_{\mu}(n)\, \delta_{n,m-\emu}+
    (I+{
    \gamma}_{\mu}){U}^{\dagger}_{\mu}(n-\emu)\, \delta_{n,m+{\emu}}.
\end{eqnarray}
The Euclidean $\gamma$ matrices in the standard representation are
defined as:
\[
{\gamma_1} \equiv  \left[ 
{\begin{array}{rrrr}
0 & 0 & 0 &  - I \\
0 & 0 &  - I & 0 \\
0 & I & 0 & 0 \\
I & 0 & 0 & 0
\end{array}}
 \right]
{\gamma_2} \equiv  \left[ 
{\begin{array}{rrrr}
0 & 0 & 0 & -1 \\
0 & 0 & 1 & 0 \\
0 & 1 & 0 & 0 \\
-1 & 0 & 0 & 0
\end{array}}
 \right] 
\]
\begin{equation}
{\gamma_3} \equiv  \left[ 
{\begin{array}{rrrr}
0 & 0 &  - I & 0 \\
0 & 0 & 0 & I \\
I & 0 & 0 & 0 \\
0 &  - I & 0 & 0
\end{array}}
 \right] 
{\gamma_4} \equiv  \left[ 
{\begin{array}{rrrr}
-1 & 0 & 0 & 0 \\
0 & -1 & 0 & 0 \\
0 & 0 & 1 & 0 \\
0 & 0 & 0 & 1
\end{array}}
 \right] .
\end{equation}%
The hermitian form of the Wilson-Dirac matrix is given by
multiplication of $M$ with $\gamma_5$:
\begin{equation} 
Q=\gamma_5\, M,
\label{HWD}
\end{equation}
with $\gamma_5$ defined as 
the product 
\begin{equation} 
\label{GAMMA5}
\gamma_5\equiv 
\gamma_1 
\gamma_2 
\gamma_3 
\gamma_4
=\left[ 
{\begin{array}{rrrr}
0 & 0 & -1 & 0 \\
0 & 0 & 0 & -1 \\
-1 & 0 & 0 & 0 \\
0 & -1 & 0 & 0
\end{array}}
 \right] .
\end{equation}

\end{appendix}

\end{document}